# Solar Physics Research in the Russian Subcontinent - Current Status and Future

Alexei A. Pevtsov[1], Yury A. Nagovitsyn[2], Andrey G. Tlatov[3], Mikhail L. Demidov[4]

[1] *National Solar Observatory, Sunspot, NM 88349, USA*

[2] *The Main (Pulkovo) Astronomical Observatory, Russian Academy of Sciences, Pulkovskoe sh. 65, St. Petersburg, 196140 Russian Federation*

[3] *Kislovodsk Solar Station of Pulkovo Observatory, PO Box 145, Gagarina Str., 100, Kislovodsk, 357700 Russian Federation*

[4] *Institute of Solar-Terrestrial Physics SB RAS, 664033, Irkutsk, P.O. Box 291, Russian Federation*

**Abstract:** Modern research in solar physics in Russia is a multifaceted endeavor, which includes multi-wavelength observations from the ground- and space-based instruments, extensive theoretical and numerical modeling studies, new instrument development, and cross-disciplinary and international research. The research is conducted at the research organizations under the auspices of the Russian Academy of Sciences and to a lesser extent, by the research groups at Universities. Here, we review the history of solar physics research in Russia, and provide an update on recent developments.

## 1 Introduction

The research in solar physics in Russia and countries that until early 1990th were part of Soviet Union has strong historical roots. Thus, for example, there are historical records of naked eyed observations of sunspots through the haze of the forest fires in 1365 and 1371 (see [1]). The Novgorod Chronicles also contain a description of about 40 solar eclipses between 1064 and 1567 over the what is now a European part of Russia and the Northern Scandinavia. The description of May 1, 1185 eclipse provides the very first known description of solar prominences: "[...] it became very dark, even the stars could be seen [...], and the sun became like a crescent of the moon, from the horns of which a glow similar to that of red-hot charcoals was emanating" [1]. With the development of astronomical studies in Russia, astronomers begun conducting the regular scientific observations of solar eclipses aiming at understanding its physics and the nature of other related phenomena (e.g., changes in the air temperature). In 1699, Russian statesman and astronomer, James Bruce (known in Russia as Yakov V. Brus, 1669 –1735) developed an instrument for observing the solar eclipses by projecting the images of the Sun on a white screen. The instrument was used by Russian czar, Peter the Great during his travel. In 1705, to prevent the general public from treating the incoming solar eclipse as a miracle, Peter the Great instructed to broadly disseminate the information about the time and location of incoming solar eclipse. Then, in his letter to Peter the Great dated 18 July 1716, Bruce writes that he observed a great number of spots on the Sun and adds that this is the first time he saw the sunspots and as far as he is aware, sunspots were not seeing for a long time. In the retrospect, these Bruce's observations were taken at the rising phase of solar cycle, and that year, the sunspot number was indeed the highest as compared with some previous years. In 1725, to promote the public eduction, Peter the Great created the Saint Petersburg Academy of Sciences, which later became the Russian Academy of Sciences, and several prominent European scientists were invited to work in Russia. Thus, for example, from 1726 till 1747, French astronomer, Joseph-Nicolas Delisle served as the director of Saint Petersburg Astronomical Observatory, where among other things he (and several of his assistants, including Christian Mayer and Georg Wolfgang

Corresponding author's email: apevtsov@nso.edu

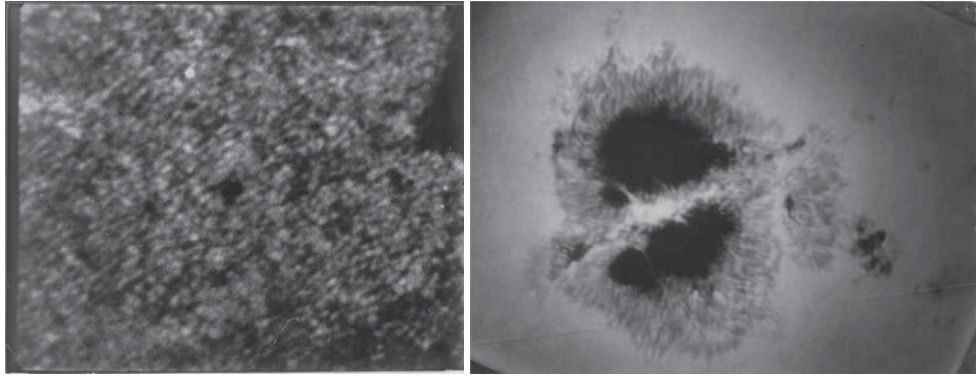

Figure 1: One of the early photographs of (left) solar granulation and (right) large sunspot by A. P. Hansky. Reproduced from the archive of the Russian Academy of Sciences at http://www.ras.ru/namorozovarchive/12 ⌐ actview.aspx?id=5256print=1

Kra*ff*t) took observations of sunspot position and studied the solar rotation. In 1738, Delisle published an article on "Theory of Sunspot Motions". The theory was based on the assumption that the solar equatorial rotation is 25.5 days, and the rotational axis of the Sun is inclined to the ecliptic plane by 7.5 degrees. The observation of sunspots from the Saint Petersburg Observatory were used for determination of solar rotation by Kra*ff*t (who then served as an assistant to Delisle at the Observatory), and Leonhard Euler (a famous mathematician, who served at the Saint Petersburg Academy of Sciences from 1727 till 1741). During these years, there were also important observations of Venus transit in 1761 by Mikhail V. Lomonosov (1711–1765) that led to the discovery of the Venus' atmosphere. In 1829, after his travel to Russia, Alexander Humboldt (1769–1859) convinced the Russia's czar to create a chain of stations for measuring geomagnetic fields (and its variations due to solar activity) across Siberia. Some of these stations later formed the basis for the modern-day research organizations.

As an introduction to the modern solar research in Russian Federation, we provide a brief overview of this early history. When it is appropriate, this overview includes solar research programs in other countries that at the time of reference were the part of Russian Empire or, later, USSR. Given the great extent of research in solar astronomy in Russia, and its interdisciplinary links to astrophysics, Earth's magnetosphere and atmosphere, laboratory plasmas, biology, and other research areas, this article can only provide a brief and somewhat general picture with many important details left out.

## 2 Early years of solar astronomy in Russia.

By late 1800th, the solar eclipses were regularly observed by Russian professional and amateur astronomers, which includes the famous Russian chemist, Dmitri Mendeleev (known for formulating the periodic table of chemical elements) who in 1887 made a solo ascent in a hot-air balloon to observe a solar eclipse above the clouded Moscow. Based on the observations of solar eclipse in 1851, Otto Wilhelm von Struve (1819 – 1905, the Director of Pulkovo Observatory and a leading member of the Russian Academy of Sciences) made the pioneering astrophysical conclusions that (1) the solar corona is the part of the Sun and is alike an atmosphere above the solar photosphere, (2) the solar prominences belong to the Sun as a celestial body, and (3) that there is an internal connection between the sunspots, plages and prominences.

The first systematic studies of solar phenomena in Russia are attributed to Fyodor A. Bredikhin (Russian: Фёдор А. Бредихин, 1831–1904) who started his regular observations of solar prominences in 1859 at the Moscow Observatory. In 1868, Kaspar Gottfried Schweizer (1816 – 1873, the Director of the Moscow Ob-servatory) purchased a spectroscope for observing the prominences outside of solar eclipses, and conducted first observations. In 1869, Schweizer prepared an article on the results of his measurements, but his illness and a premature death prevented this publication. After Bredikhin replaced Schweizer as the next director of Moscow observatory, he continued the observations of prominences for about 11 years from 1873–1884. Between 1881–1910, observations of prominences were also conducted by A. K. Kononovich in the Odessa Astronomical Observatory (Ukraine). Thus, the spectroscopic studies of solar prominences and their long-term synoptic observations in Russia begun at about the same time as in other European observatories.

In 1865, the photographic solar patrol was started by Russian astronomer Matvey Gusev (1826–1866) in Vilnius Observatory (Lithuania), and about 900 photographs of the Sun were taken between 1868 and 1876. Unfortunately, in 1876, the Vilnius observatory suffered a major damage due to a fire, and in 1881, it was closed [2]. Some historians suggest that about 800 photographs of the Sun taken in Vilnius observatory were sent to the Pulkovo observatory. However, there are no records of these data in the archive of the Main (Pulkovo) Astronomical Observatory.

Beginning 1884, regular sunspot drawings were taken with the 6-inch telescope at the Tashken Astronomical Observatory (TAO, now Ulugh Beg Astronomical Institute of the Uzbek Academy of Sciences, Uzbekistan); the drawings were supposedly sent to Zürich [3]. Beginning 1936, the TAO took part in the international flare patrol program conduced at the suggestion of G.E. Hale under the auspices of the International Astronomical Union [3]. In 1894–1905, the observations of solar corona, prominences and the photographic observations of solar corona, granulation (Figure 1) and sunspots were conducted in Pulkovo Observatory by Alexei P. Hansky (Russian: Алексей П. Ганский, 1870–1908). Hansky took park in several expeditions to observe solar eclipses. He climbed Mon Blanc (the highest peak in Europe outside of the Caucasus mountain range) nine times to take astronomical observations of solar constant, to observe solar corona and planet Venus. In 1897, he discovered that the shape of solar corona changes with the solar cycle. Hansky had a passion for photography, and he succeeded in taking photographs of sunspots and granulation of extremely high quality, that were exceeded only much later by observations from telescopes on the stratospheric balloons. Hansky used a high cadence photographic observations to estimate the lifetime of the photospheric granules as 2-5 minutes. The photographic observations of sunspots and studies of solar rotation (using sunspots as tracers) were also conducted at the Moscow Observatory by Aristarch A. Belopolsky (Russian: Аристарх А. Белопольский, 1854–1934), who was a student of F.A. Bredikhin [4]. In 1888, Belopolsky joined the Pulkovo Observatory near Saint Petersburg, where he continued studies of solar rotations using spectroscopic observations. He discovered that the rotation of plages follows the same profile of rotation as the sunspots and found a small decrease in solar equatorial rotation rate during 1925–1933, which he contributed to possible solar cycle variations. In 1915, Belopolsky estimated the temperature of sunspots on the basis of spectrophotometic measurements (the method was proposed earlier by a famous Russian physicist, Pyotr Lebedev, 1866–1912). Other notable Russian astronomers who conducted solar research in the late 19th- early 20th century were V.K. Tserasky (1849–1925, solar photometry), B.P. Gerasimovich (1889–1937, solar corona), G.A. Tikhov (1875–1960, solar eclipses), Y.Y. Perepyolkin (1906 – 1940, solar rotation, formation height of the chromosphric plages, the physics of the prominences, and the structure of the chromosphere). In addition to Moskow and Pulkovo observatories, by early 20th century, solar observations were also conducted in Kharkov Astronomical Observatory (Ukraine), Tashkent Astronomical Observatory, Abastumani Astronomical Observatory (Georgia, [5]) and several other smaller observatories.

The need for a broad international collaboration in solar studies was recognized by many astronomers and in mid-1904, academies/science societies of seven countries (including Russia) agreed to form the national commissions for cooperation on solar research; it was also agreed to send the representatives of these national commissions to the Conference on Solar Research in St. Louis (USA) in conjunction with the International Congress of Arts and Science (see, copy of the letter by G.E. Hale in archives of the Russian Academy of Sciences at http://www.ras.ru//MArchive/pageimages/543%5C11_076/001.jpg and http://www.ras.ru//MArchive/pageimages/543%5C11_076//002.jpg). The delegates of this conference, made the decision to create the International Union for Cooperation in Solar Research, which was one of the precursor organizations of the International Astronomical Union (formed in 1919). About four months after the meeting in St. Louise, the Director of Pulkovo Observatory, Oskar Backlund (1846 – 1916) initiated the creation of Russian Division of this International Union. At the first meeting of a newly created division, Hansky put forward a proposal to build a high-altitude heliophysical observatory in the southern part of Russia (Pamir mountains) and in cooperation with Mount Wilson observatory in USA conduct a near continuous observations of the Sun. In years following the October revolution in 1917, the Russian Division of the International Union was dissolved, and in 1930, it was replaced by a newly created Commission on Study of the Sun. At the first meeting, the commission discussed creation of a new high altitude astronomical observatory. In 1932, this new Abastumani Astronomical Observatory (Georgia) was created by Eugene Kharadze. Also, in 1932, upon a suggestion of Perepyolkin, the commission recommended creation of a network of synoptic solar observations, named "Solar Service" (Russian: Служба Солнца [see 6]). Initially, the observations were taken in three observatories: Tashkent Astronomical Observatory, Kharkov Astronomical Observatory, and the Main (Pulkovo) Astronomical Observatory [7]. In 1937, the first catalog of solar activity covering 1933–1937 observations was compiled by Pulkovo astronomers Mstislav N. Gnevyshev (1914–1992) and B.M Rubashev. Catalogs of solar activity for later periods (1938–1992) were compiled by Raisa S. Gnevysheva. The need for synoptic observations was in part justified by early studies by several scientists including Alexander L. Chizhevsky (Russian: Александр Л. Чижевский) of the effects of solar activity cycles on Earth's climate and human activities.

As the rest of the Russian astronomical community, solar physics was devastated by the Stalinist purges in 1936–1937 (about 30% of Russian astronomical community were imprisoned or executed, [8]), and Pulkovo Observatory was hit especially hard. To make things worse, during the WWII, several major observatories in occupied regions of Soviet Union including the Pulkovo Observatory were largely destroyed; some astronomical institutions were evacuated from the European part of Russia and continued limited operations from newly established locations. Thus, for example, in 1942, astronomers from the Sternberg Astronomical Institute of Moscow State University and the astronomical observatory of Kiev University established a new department at Sverdlovsk State University, where they provided the daily forecasts of "radio-weather" for the needs of radio-communications and army radio-intelligence services. The group also calculated the time of sunrise and sunset for the long-range bombing operations. Éval'd Mustel' (1911 – 1988) led the work of this group.

Biographical sketches of the above astronomers can be found at [9].

## 3   Post-World War II – mid-1990th

After the end of WWII, the solar astronomy in Soviet Union experienced significant growth. The Main (Pulkovo) Astronomical Observatory was rebuilt during 1945–1954. In October 1948, first photoheliograms were taken from the new high altitude astronomical station near Kislovodsk, which later became known as the Kislovodsk Mountain Astronomical Station (KMAS). The station was founded by M.N. Gnevyshev. In 1954, Pulkovo Observatory begun regular (monthly) publication of summaries of solar activity "Solnechnye Dannye" (Solar Data, Russian: Солнечные Данные). Shortly after the end of war, new solar instruments were built in Kislovodsk Mountain Astronomical Station (coronagraph and photoheliograph), the Crimean Astrophysical Observatory (coronagraph, 1950; the large tower solar telescope, 1954), Abastumani Observatory (photoheliograph, 1950), and Pulkovo Observatory (horizontal solar telescope, 1951). In later years, number of solar instruments (e.g., H$\alpha$ telescopes, photoheliographs, horizontal telescopes, coronagraphs etc) were built in many other observatories across Soviet Union (e.g., the Central Astronomical Observatory in Kiev, Ukraine, Alma Ata station (now, Tien Shan Astronomical Observatory, Kazakhstan), Shamakhy Astrophysical Observatory (Azerbaijan, founded in 1960), Astronomical Observatory of Ural State University (founded in 1965), Astronomical Institute of Uzbek Academy of Sciences (founded in 1873 as Tashkent Astronomical Observatory), Sayan Solar Observatory (founded in 1964), and the Baikal Astrophysical Observatory of the

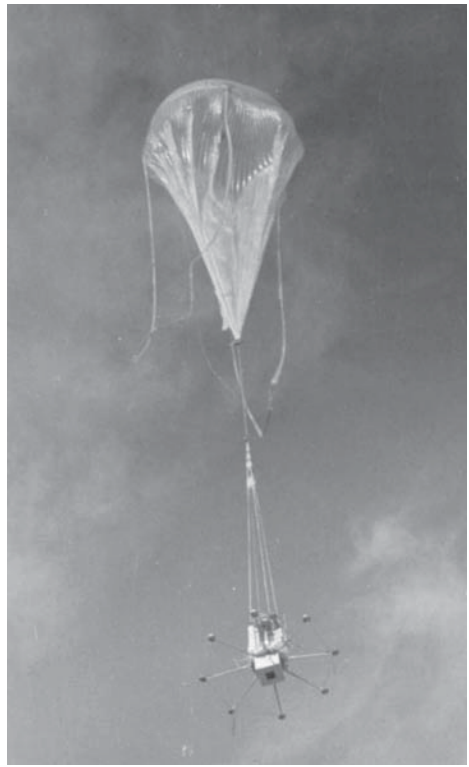

Figure 2:The soviet stratospheric solar telescope in flight.

Institute of Solar-Terrestrial Physics, and Ussuriysk Astrophysical Observatory (founded in 1953)). After the first soviet radio observations during the solar eclipse in 1947, the solar radio astronomy was also rapidly developing with the most notable radiotelescopes such as Big Pulkovo Radio Telescope, [10], RATAN-600 (600 meter diameter ring-telescope in Northern Caucasus, [11]), Siberian Solar Radio Telescope (SSRT, Irkutsk, [12]), and Pushchino Radio Astronomy Observatory.

Mid-1960–late 1970th were the heyday of the solar astronomy in Soviet Union. In 1960, the Siberian Institute of Terrestrial Magnetism, Ionosphere, and Radiowave propagations (SibIZMIR, in 1992, renamed to Institute of Solar-Terrestrial Physics) was created on the basis of Irkutsk magneto-ionospheric station. In early-mid 1960th, several observatories started development of line-of-sight (LOS) and vector magnetographs (Institute of Terrestrial Magnetism, Ionosphere, and Radiowave Propagation, IZMIRAN; the Crimean Astrophysical Observatory, CrAO; the Main (Pulkovo) Astronomical Observatory, GAO; SibIZMIR/ISTP). The magnetic field measurements were measured on the basis of Zeeman effect (photospheric and chromospheric spectral line) as well as Hanle effect (chromospheric prominences). The theory of the polarized light radiative transfer was developed by V.E. Stepanov (CrAO and later, SibIZMIR) and D. Rachkovsky (CrAO). At the end of 1960th, the world-best vector magnetograph was operating at Sayan Solar Observatory operated by SibIZMIR. The observations from this magnetograph led to discovery of a crossover effect in sunspots [13]. SibIZMIR/ISTP also built small magnetographs for observatories in East Germany and Czechoslovakia.

In 1957-58, the network of solar stations in USSR took active participation in the International Geophysical Year (IGY), and in 1964-65, the International Quiet Sun Year coordinated programs. IGY culminated with the launch of Sputnik in 4 October 1957, the Earth first artificial satellite. Two years later, the first ever direct observations and measurements of the solar wind were conducted from the Luna-1 ("Dream", Russian: Мечта). In 1966, the first Soviet solar telescope (project "Saturn", Figure 2) was flown on a stratospheric balloon (total 4 flights in 1 Nov. 1966, 22 Sept 1967, 30 Jul 1970, and 20 Jun 1973, first three flight were with 0.5 m telescope, the last flight was with 1 m telescope). The last flight produced images with 0.12 arcsecond resolution [14]. From 1977–1989, the high resolution observations were continued using a telescope at Eastern Pamir mountains at altitude of 4,330 meters [15]. In 1980, the Large Solar Vacuum Telescope (LSVT) was built at the Baikal Astrophysical Observatory [16], and in late 1980th, a country-wide site survey was conducted for a future high resolution solar telescope.

From December 1974 till late–1976, spectral observations of solar flares, flocculi and prominences were taken with far-EUV spectrometer (97–140 nm) on board the Salyut–4 space station (Orbiting Solar Telescope, OST-1, 25 cm aperture and 2.5 focal length of 2.5 m, [17]). Studies of solar radio, EUV, X-ray and γ-ray fluxes from the Sun, properties of solar wind and the effects of solar activity of Earth's magnetosphere and ionosphere were studies from sounding rockets and Earth-orbiting satellites launched in the framework of "Interkosmos" international program, and on-board of specialized satellites series "Prognoz". "Intercosmos" program (26 launches) operated from 1970 till mid-1990th was a close collaboration between USSR and Eastern European countries; other participating countries included Afghanistan, Cuba, France, India, Mongolia, Syria, and Vietnam. Twelve satellites of "Prognoz" series were launched between 1972–1996 (Prognoz-11 and Prognoz-12 were named Interbol–1 and Interbol–2). Instruments for heliophysics studies of the sun, solar wind, Earth's magnetosphere and upper atmosphere were also launched in numerous spacecrafts of "Kosmos" and "Electron" series.

By mid-1960th the "Solar Service" network had rapidly grown to include 20 observatories located across eleven time zones in USSR, and it even included a horizontal solar telescope in Cuba [18]. The network stations were equipped with near identical telescopes to take daily observations of the Sun in the photosphere and chromosphere; some stations were also taking coronagraphic and radio observations. The backbone of the "Solar Service" was formed by the Main (Pulkovo) Astrophysical Observatory and its Kislovodsk Mountain Astronomical Station, Crimean Astrophysical Observatory (CrAO), Shamakhy Astrophysical Observatory (Azerbaijan, Astronomical Observatory of Ural State University, Astronomical Institute of Uzbek Academy of Sciences, Sayan Solar Observatory of the Institute of Solar-Terrestrial Physics, and Ussuriysk

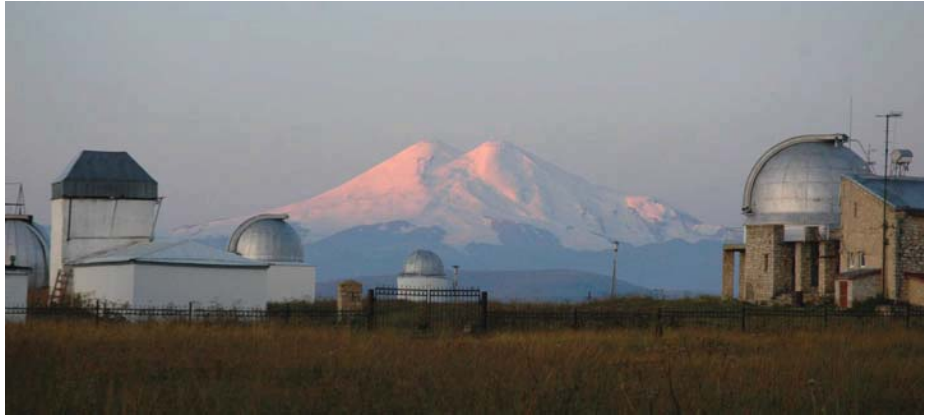

Figure 3: View of Kislovodsk Mountain Astronomical Station. The double-peaked mountain is Mount Elbrus, Europe's highest mountain (5,642 m).

Astrophysical Observatory. Sever decline in funding that followed the disintegration of Soviet Union resulted in discontinuation of "Sun Service" operations in about 1997. Some observatories (e.g., KMAS, ISTP, CrAO) continued synoptic operations despite the funding shortage. The significant funding cuts in early-1990th had the devastating consequences on solar physics research in Russia. The plans for a new high resolution solar telescope were canceled, new instrument development was put on hold, and the operations of many observatories were reduced. Solar observations from space continued with a different degree of success. For example, in early 1990th, the idea for a stereoscopic observations of the Sun from Lagrangian L4 and L5 points was put forward [19], but the mission was not funded due to the budget cuts. X-ray telescope on board Fobos–1 (Phobos–1) spacecraft, launched in 1988 returned about 140 images during its cruising to Mars, [20]. Koronas-I and Koronas-F satellites, launched in 1994 and 2001 had a suite of EUV, X-ray and radio telescope instruments for studying solar flares [21]. Its follower, Koronas-Foton, was launched in January 2009 and successfully operated until December 2009, when the communications with the satellite were lost.

## 4   Modern-days solar ground-based research.

Currently, the solar research in Russia is conducted in more than 20 organizations under the auspices of section "The Sun" of the Scientific Council on Astronomy of the Russian Academy of Sciences. The research conducted in these organizations covers all area of solar physics and includes theory, numerical modeling, instrument development, and observations. Here we address only the research directly related to study of solar phenomena. Many organizations (including those reviewed below) conduct magnetospheric and ionospheric research (including cosmic rays); while such research is relevant to the understanding of the effects of solar activity on our planet, it is not included in this review due to the article's length limitations.

The Main (Pulkovo) Astronomical Observatory (Saint Petersburg, www.gao.spb.ru/english/) and its Kislovodsk Mountain Astronomical Station (en.solarstation.ru/, Figure 3) conduct a vigorous program of synoptic observations. On average, there are 334 sunny days at KMAS. Daily observations are taken in the photosphere (full disk photoheliograms, broadband at about 4200 Å) and chromosphere (Hα, 10 cm refractor, Halle filter with 0.25 Å bandpass). Photospheric observations are used for determination of daily sunspot number and group and sunspot areas. Observations with small (20 cm aperture) Lyot-type coronagraph are taken in two spectral lines (Fe XIV(green) 5303Å, and Fe X (red) 6374 Å) since 1957. Daily sun-integrated radio-flux in 5 cm (6150 MHz) is observed with a 3-meter parabolic antenna. Radio observations were conducted from mid-1960th till late 1994, and then restarted in 2002. In addition to solar observations, the two components of geomagnetic field are measured with Simple Aurora Monitor (SAM) magnetometer. In addition to the scientific observations, KMAS is used for a summer practice of undergraduate students from the Saint Petersburg State University as well as several local universities. Figure 4 provides example of graphic summary of KMAS daily observations. The chromospheric data are used to construct the synoptic maps of polarity of large scale magnetic fields and neutral lines. Observations of solar chromosphere in Ca K II spectral line are taken with spectroheliograph (in operation since 1960), and the new Solar Patrol Optical Telescope (SPOT). Figure 5 (two right panels) show example of spectra taken with SPOT and full disk images created from these spectra at a selected *off*set in wavelength. SPOT has been developed as a prototype of the automatic Hα and Ca II K line telescope–spectroheliograph for small university observatories. The current plans include creation of a network of these telescopes with instruments located at the Main (Pulkovo) Astronomical Observatory, Crimean Astrophysical Observatory, Observatory of Ural State University, Baikal Astrophysical Observatory, and Ussuriysk Astrophysical Observatory. To complete the global coverage, additional sites in Cuba and Bolivia are also considered.

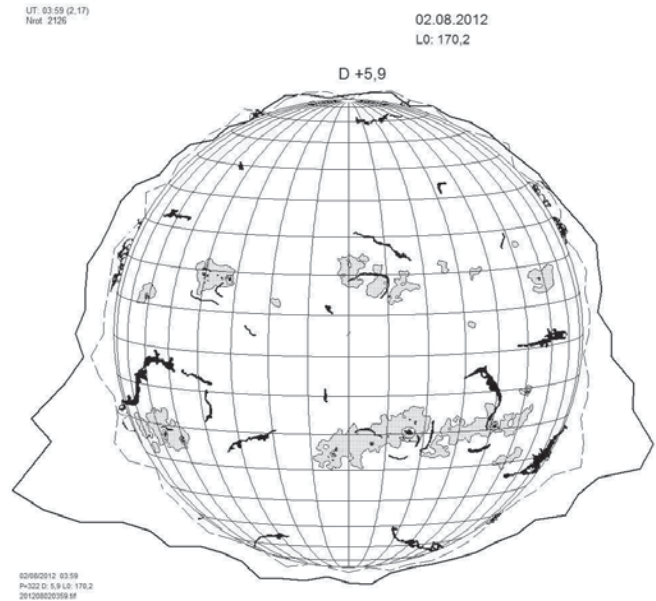

Figure 4: Example of summary of daily observations from KMAS showing sunspots (dark and the chromospheric plages (closed dotted areas on the disk), the filaments and prominences (black elongated features on the disk and *off* limb), the outer extent of the solar corona in Fe XIV(green) 5303Å (solid line outside solar disk), and Fe X (red) 6374 Å (dashed line).

Scientific research in Pulkovo Observatory (including KMAS) concentrates on developing a theoretical understanding of physics of (eruptive and non-eruptive) solar phenomena in solar photosphere, chromosphere and corona (including oscillations) as well as the observational studies of these phenomena. In addition, the research includes interplanetary medium and propagation of CME shocks. Other direction of research aims at understanding the solar cycle, including solar and stellar dynamos, anomalies of recent cyclic activity, reconstruction of past solar cycles, and the *effects* of solar activity on Earth climate. Research pertinent to space weather utilizes observations from KMAS as well as other ground- and space- based instruments.

Institute of Solar-Terrestrial Physics (ISTP, Irkutsk, en.iszf.irk.ru/Main Page) operates two optical observatories (Baikal, en.iszf.irk.ru/Baikal_Astrophysical_Observatory and Sayan, en.iszf.irk.ru/Sayan Solar Observatory) that conduct the research in solar physics in optical wavelength range and one major radio facility, the Siberian Solar Radio Telescope (SSRT, en.iszf.irk.ru/The_Siberian_Solar_Radio_Telescope). Two full disk solar telescopes (Hα and Ca II K line) operate at Baikal Observatory, but instead of synoptic observations, the emphasis is made on case studies by the individual ISTP researchers. Hα are taken with a Halle filter with 0.5 Å bandpass, tunable to ± 1 Å from the line center. The Large Solar Vacuum Telescope (LSVT, 1 meter heliostat mirror, Figure 7) constructed in early 1980 is undergone recent renovations. LSVT is now used for full Stokes spectropolarimetry with high spatial resolution in solar flares. ISTP has a strong heritage in developing and operating of solar LOS and vector magnetographs (mid-1960 – later 1980th). In 2014, a new full solar disk vector magnetograph was developed and now is in the commissioning stage. The new instrument, the Solar Synoptic Telescope (SOLSYT), will be installed in a new dome at the Baikal Observatory. The telescope was manufactured and delivered to the ISTP. Initial tests had verified the instrument performance in respect to polarization measurements. ISTP's Sayan Solar Observatory operates the 800-mm aperture horizontal solar telescope equipped with a set of magnetographs and spectrophotometers; a specialized telescope designed to observe the magnetic fields of the Sun as a star and to measure weak, large-scale magnetic fields on the solar surface, and 530-mm aperture coronagraph. As a more recent development, the ISTP initiated a development of a new network of full disk low resolution LOS magnetographs. The instruments are based on a previous design of the Solar Telescope of Operational Prognosis (STOP, [22]). The network consists of three magnetographs now in operation at the Ussuriysk Astrophysical Observatory, Baikal Astrophysical Observatory and the Kislovodsk Mountain Astronomical Station. The data are used for modeling of large scale magnetic topology of solar corona, and space weather modeling and forecast. The advantage of STOP magnetographs as compared with other instruments (e.g., magnetograph at Wilcox Solar Observatory) is a possibility to observe in many spectral lines simultaneously. This enables some new tasks, connected e.g. with interpretation of Stokespolarimeter measurements of the quiet Sun magnetic fields [23]. Figure 6 shows examples of synoptic radial magnetogram observed at KMAS, and the charts of the derived parameters.

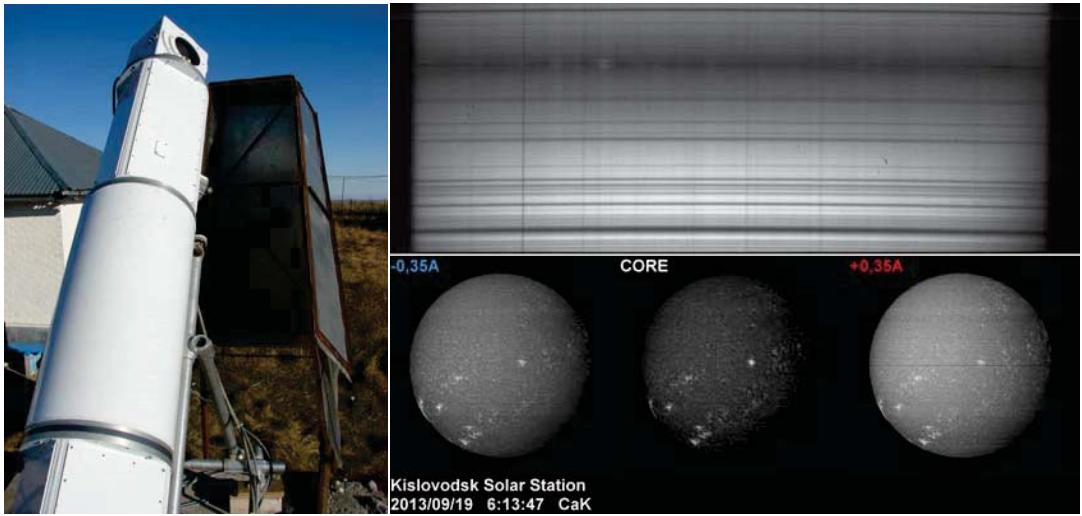

Figure 5: The Solar Patrol Optical Telescope (left) and example of data taken with it in Ca II K chromospheric spectral line (spectra – upper right, and images created from spectra – low right).

The Siberian Solar Radio Telescope (SSRT) is a radio facility for solar synoptic observations built and operated by the ISTP. It is located about 220 km from the city of Irkutsk. Designed as a crossed interferometer, it consists of two arrays of 256 (128 x 2) parabolic antennas 2.5 meters in diameter each, spaced equidistantly at 4.9 meters and oriented in the E-W and N-S directions (see photo in [24]). The length of each linear baselines of the interferometer is 622.3 meters; the instrument's working wavelength is $\lambda = 5.2$ cm (5.7 GHz). The SSRT was commissioned in 1984, and its regular synoptic observations started in 1998 [12].

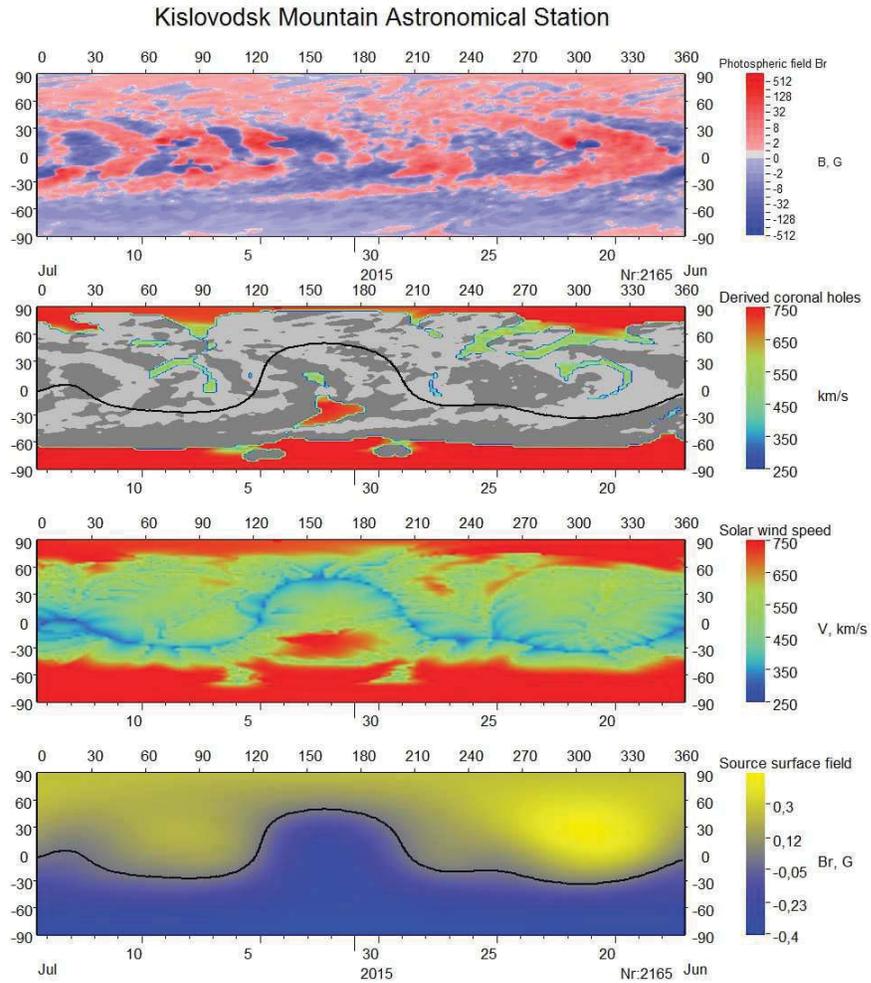

Figure 6: Example of synoptic maps of radial magnetic field observed by STOP (upper panel), and the synoptic maps of areas of open field lines (coronal holes), solar wind speed, and the source surface magnetic field derived from the magnetic field data.

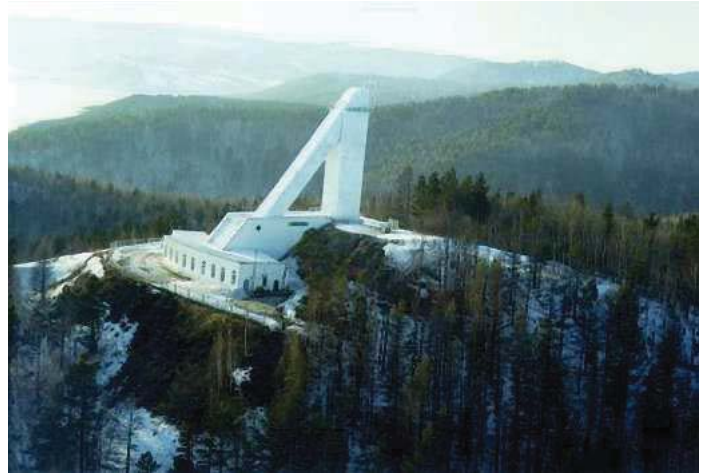

Figure 7: The aerial view of the Large Solar Vacuum Telescope at Baikal Astrophysics Observatory on the northern shore of Lake Baikal.

ISTP scientists conduct the research in a broad area of topics ranging from solar dynamo, solar magnetism on different spatial and temporal scales, the observational spectroscpcoy and spectropolarimetry, solar oscillations in the photosphere and chromosphere, and theoretical and numerical modeling of solar eruptive and non-eruptive processes, particle acceleration, and evolution of interplanetary CMEs. The main directions of research at ISTP solar department are: (1) the nature of solar variability, and the role of magnetic fields (large-scale and small scale ones) in solar activity; (2) the physical conditions in sunspots; (3) spatial structure and dynamical processes in photosphere, chromosphere, corona; (4) oscillations and waves in solar atmosphere on different levels; and (5) development of new instruments (solar telescopes equipped with polarimeters, narrow-band filters, control systems of telescopes).

Solar research at the Institute of Terrestrial Magnetism, Ionosphere, and Radiowave Propagation (IZMIRAN, Moscow, http://www.izmiran.ru/?LANG=en) focuses on physics of solar phenomena and the solar-terrestrial effects of solar activity. This includes the study of the Sun as the driving source of space weather, the propagation of solar disturbances from the Sun to Earth, solar wind and its interaction with the Earth's magnetosphere, the effects of solar activity on Earth's upper and middle atmosphere, ionosphere-magnetosphere coupling, the effects of solar activity on Earth's weather, climate and the technological systems. The IZMIRAN's Solar Radio Laboratory operates a radiospectrograph (frequency range 25 - 270 MHz, time resolution – 0.04 s/0.02 s alert mode, frequency resolution – 0.2 MHz), and fixed frequencies radiometers (frequencies: 169 MHz, 204 MHz, 3000 MHz, time resolution – 1 s). Daily observations are taken betweem 0600 UT and 1200 UT (0700 – 1200 UT in winter). IZMIRAN also compiles the forecast summaries of solar and geomagnetic activity using the materials from the Space Weather Prediction Center (SWPC, Boulder, Colorado, USA), the Solar Influences Data Aanalysis Centre (SIDC, Brussels, Belguim), and its own data.

Solar group at Special Astrophysical Observatory (SAO) operates the RATAN-600 radiotelescope (http://www.sao.ru/hq/sun/index.html) situated in the southern part of European part of Russia, the Northern Caucasus. The telescope antenna has a diameter of 567 meters, and is formed by 895 independent elements 2 by 11.5 meters in size. The orientation of antenna elements is controlled by a computer to create the shape of outer ring of a parabolic mirror with 567 meter in diameter. The antenna is divided on four sectors (southern, northern, eastern, and western) with 225 antenna elements in each sector. Up to three sectors can employed to conduct simultaneous but independent observations. For that observing mode, there are three mobile observing stations with secondary mirrors and the electronics for recording the data. These mobile stations can be placed at 12 fixed azimuthal positions. Daily observations of the Sun are taken at about 0900 UT. Typical observations include one dimensional scans in 30-40 wavelengths in the range 1.67 – 32 cm in both left- (LCP) and right- (RCP) circular polarization. The observations are used for routine measurements of magnetic field is solar corona at two different heights, the studies of solar flares, development of space-weather forecasting capabilities using multi-frequency radio observations, and developing methods for measuring the plasma properties throughout the solar atmosphere. Studies of the Sun in radio wavelengths conducted at the Radiophysical Research Institute (NIRFI, Nizhny Novgorod, Russian Federation) concentrate on the physics of solar energetic events, monitoring of solar radio flux, flare forecast, and the effects of solar activity on processes in Earth's atmosphere and interplanetary space.

The solar physics research at the Institute of Applied Physics (http://www.ipfran.ru/) focuses on theoretical and observational studies of origin of the Sun's electromagnetic emission in different frequency/wavelength ranges, coronal heating, the particle acceleration in flares, the structure and dynamics of solar active regions in the corona, and the development of kinetic theory of the solar wind.

The laboratory of X-ray astronomy of the Sun at the Lebedev Physical Institute (aka the Physical Institute of the Academy of Sciences, or FIAN) is one of the leading centers of solar research in Russia. The laboratory conducts both theoretical and observational research on structure and dynamics of solar corona, and the mechanisms of energy release in the solar atmosphere (flares and the chromospheric/coronal heating). understanding the mechanisms for coronal heating and the processes of the photospheric heating in flare kernels, the particle acceleration in flares, the structure and dynamics of various X-ray phenomena and the corresponding magnetic fields (e.g., solar active regions, coronal bright points, coronal holes), and the understanding of mechanisms responsible for the solar wind acceleration. The laboratory leads the development of solar X-ray telescopes and spectrometers for the experiments in space. Figure 8 shows example of images taken with telescopes on board Koronas-Foton in August 2009. These instruments were part of TESIS (http://www.tesis.lebedev.ru/en/about tesis.html) suite of instruments developed at the FIAN.

Laboratory of Solar Physics of Crimean Astrophysical Observatory (craocrimea.ru, the website is in Russian only) operates two tower telescopes (BST-1 and BST-2), and two coronagraphs. The group also conducts a synoptic program that includes daily observations of the Sun in He I 1083.0 nm and Hα, magnetographic observations of the global magnetic field, visual measurements of magnetic field in sunspots, and sunspot number. The 50-cm aperture tower telescope BST-1 is used to take spectral and magnetic field observations of different solar features both with high spatial resolution and in sun-as-a-star mode.The BST-1 is equipped with a spectroheliograph designed to take full disk spectroheliograms simultaneously in two spectral lines (e.g., Hα and Ca II K line). The telescope is used to measuring the global magnetic field of the sun, and the solar global oscillations. The BST-2 is equipped with two main mirrors of 20-cm and 45-cm in diameter, and it is used for taking daily measurements of sunspot field strengths, and full disk images of the Sun in He I 1083.0 nm. The telescope is also used for spectral observations in flares, plages, coronal holes, and quiet Sun areas in the wavelength range between 390 nm and 1100 nm. Data collected from these instruments are used for studying the physics of various solar features, solar magnetic field on different spatial and temporal scales, solar eruptive events, and solar-terrestrial connections.

Research on dynamics and spatial distribution of solar activity is conducted at the Ussuriysk Astrophysical Observatory in Russia's Far East (http://www.uafo.ru/, the website is in Russian only). The solar physics group at the observatory also studies the properties of solar magnetic fields and their connection with other active phenomena on the Sun. The research also include the effects of solar activity on interplanetary space and Earth's atmosphere (space weather). The observatory operates 530mm aperture coronagraph, 440 mm aperture horizontal solar telescope ASU-5, 8-meter aperture RT- 8 radiotelescope (600, 750, 1000 MHz), and 150-mm aperture refractor for white light images of the Sun.

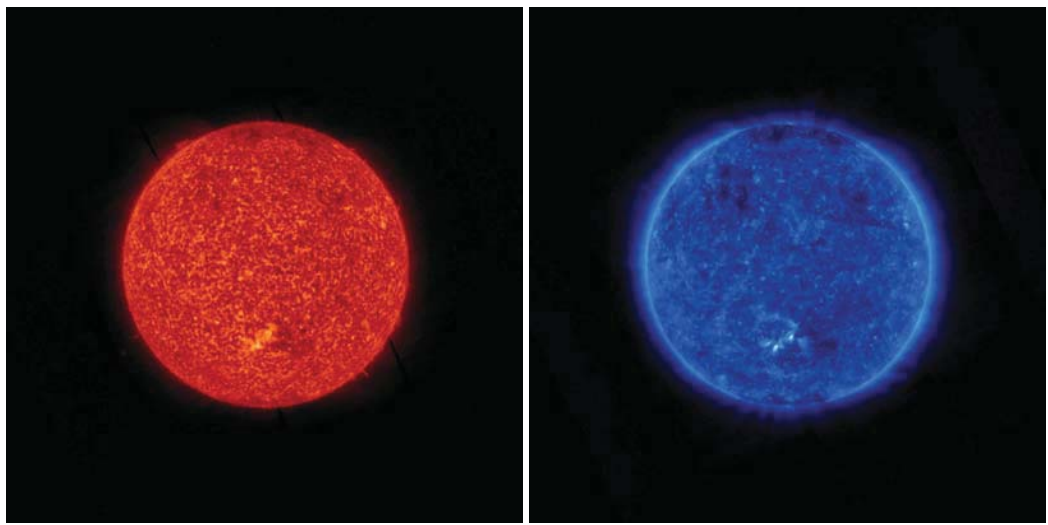

Figure 8: Example of images taken from KORONAS-FOTON spacecraft in August 2009, in (left) He II 304 Å, and (right) in Fe IX 171 Å.

In addition to research centers and observatories operated by the Russian Academy of Sciences, the research in solar physics is also conducted by the university researchers. The major groups operate at the Moscow State University (MGU), Saint Petersburg State University (SPbU), and the Ural State University. Solar physics research at Moscow State University and its Sternberg Astronomical Institute concentrates primarily on studies of magnetic reconnection, particle acceleration and flare related kinetic and MHD processes in solar atmosphere. Research on magnetoconvection, dynamo and solar/stellar cycles is conducted by individual researchers in other MGU departments. Observations from ground- and spacebased instruments (e.g., Hinode, SDO) are used. Heliophysics group at V. V. Sobolev Astronomical Institute of Saint Petersburg State University conducts most of its research in a mm and cm radiowavelength range. The current studies include the investigations of (1) the structure, evolution and magnetic fields solar prominences; (2) the spectropolarimetric and temporal structure of microwave bursts aimed at understanding the physics of these processes, the energy release and the particle acceleration, and (3) the large-scale structures in the chromosphere. The group uses the observations from the Russia's largest microwave Bauman's radiotelescope RT-7.5 operated by the Bauman Moscow State Technical University, RATAN-600, and several spacebased instruments (e.g., Hinode, SDO). Research group at the Ural State University operates a horizontal solar telescopes at the university's Kourovka Astronomical Observatory (http://astro.ins.urfu.ru/en/node/246). The data are used for study magnetic properties of active regions in solar cycle, and properties of prominences in Hα and Ca II K and H spectral lines.

In July 2015, the Russian Academy of Sciences announced plans for creation of the national heliophysical complex (nicknamed "mega-project") for continuous monitoring of solar activity and space weather forecast. The complex will included the multi-wavelength radioheliograph for all-weather monitoring of solar activity, the system of new generation radars and the large-aperture (3 meter) solar telescope with the coronagraphic capabilities. The system of four coherent scatter radars will be places at different longitudes across Russia to supplement the international Super Dual Auroral Radar Network (SuperDARN). The complex will include one incoherent scatter radar for studying ionosphere and upper atmosphere. To provide an extended coverage for studying solar activity, two geographically separated redioheliographs will be constructed: one near the current location of SSRT, in Eastern Siberia, and the other near the RATAN-600 in the Northern Caucasus. At each location the radiotelescopes will include three separate T-shaped grid with 1-km in length, and each grid will be comprised of about one hundred dishes of 3-meter dishes, two hundred dishes of 1.8 meter in diameter, and four hundred 1-meter dishes to enable the solar observations in multiple frequencies. One of the goals of this complex is to re-build the new "Solar service" program in Russia and the neighboring countries. Solar studies from space-based instruments include the Interhelioprobe mission (launch is now planned for 2020).

**Acknowledgments:** The National Solar Observatory (NSO) is operated by the Association of Universities for Research in Astronomy, AURA Inc under cooperative agreement with the National Science Foundation (NSF).